

\font\titlefont = cmr10 scaled\magstep 4
\font\sectionfont = cmr10
\font\littlefont = cmr5
\font\eightrm = cmr8

\def\sss{\scriptscriptstyle}

\newcount\tcflag
\tcflag = 0  
\ifnum\tcflag = 0 \magnification = 1200 \fi  

\global\baselineskip = 1.2\baselineskip
\global\parskip = 4pt plus 0.3pt
\global\abovedisplayskip = 18pt plus3pt minus9pt
\global\belowdisplayskip = 18pt plus3pt minus9pt
\global\abovedisplayshortskip = 6pt plus3pt
\global\belowdisplayshortskip = 6pt plus3pt

\def\barsoff{\overfullrule=0pt}


\def\endignore{}
\def\ignore #1\endignore{}

\newcount\dflag
\dflag = 0


\def\monthname{\ifcase\month
\or January \or February \or March \or April \or May \or June%
\or July \or August \or September \or October \or November %
\or December
\fi}

\newcount\dummy
\newcount\minute  
\newcount\hour
\newcount\localtime
\newcount\localday
\localtime = \time
\localday = \day

\def\advanceclock#1#2{ 
\dummy = #1
\multiply\dummy by 60
\advance\dummy by #2
\advance\localtime by \dummy
\ifnum\localtime > 1440 
\advance\localtime by -1440
\advance\localday by 1
\fi}

\def\settime{{\dummy = \localtime%
\divide\dummy by 60%
\hour = \dummy
\minute = \localtime%
\multiply\dummy by 60%
\advance\minute by -\dummy
\ifnum\minute < 10
\xdef\spacer{0} 
\else \xdef\spacer{}
\fi %
\ifnum\hour < 12
\xdef\ampm{a.m.} 
\else
\xdef\ampm{p.m.} 
\advance\hour by -12 %
\fi %
\ifnum\hour = 0 \hour = 12 \fi
\xdef\timestring{\number\hour : \spacer \number\minute%
\thinspace \ampm}}}



\def\endtitle{}
\def\title#1\endtitle{\vskip.5in\titlefont
\global\baselineskip = 2\baselineskip
#1\vskip.4in
\baselineskip = 0.5\baselineskip\rm}

\def\endauthors{}
\def\authors#1\endauthors{#1}

\def\endabstract{}
\def\abstract#1\endabstract{\vskip .3in%
\centerline{\sectionfont\bf Abstract}%
\vskip .1in
\noindent#1}

\newcount\nsection
\newcount\nsubsection

\def\section#1{\global\advance\nsection by 1
\nsubsection=0
\bigskip\noindent\centerline{\sectionfont \bf \number\nsection.\ #1}
\bigskip\rm\nobreak}

\def\subsection#1{\global\advance\nsubsection by 1
\bigskip\noindent\sectionfont \sl \number\nsection.\number\nsubsection)\
#1\bigskip\rm\nobreak}

\def\topic#1{{\medskip\noindent $\bullet$ \it #1:}}
\def\endtopic{\medskip}

\def\appendix#1#2{\bigskip\noindent%
\centerline{\sectionfont \bf Appendix #1.\ #2}
\bigskip\rm\nobreak}


\newcount\nref
\global\nref = 1

\def\ref#1#2{\xdef #1{[\number\nref]}
\ifnum\nref = 1\global\xdef\therefs{\noindent[\number\nref] #2\ }
\else
\global\xdef\oldrefs{\therefs}
\global\xdef\therefs{\oldrefs\vskip.1in\noindent[\number\nref] #2\ }%
\fi%
\global\advance\nref by 1
}

\def\listrefs{\vfill\eject\section{References}\therefs}


\newcount\nfoot
\global\nfoot = 1

\def\foot#1#2{\xdef #1{(\number\nfoot)}
\footnote{${}^{\number\nfoot}$}{\eightrm #2}
\global\advance\nfoot by 1
}


\newcount\nfig
\global\nfig = 1

\def\fig#1{\xdef #1{(\number\nfig)}
\global\advance\nfig by 1
}


\newcount\cflag
\newcount\nequation
\global\nequation = 1
\def\eqlabel{(1)}

\def\nexteqno{\ifnum\cflag = 0
\global\advance\nequation by 1
\fi
\global\cflag = 0
\xdef\eqlabel{(\number\nequation)}}

\def\lasteqno{\global\advance\nequation by -1
\xdef\eqlabel{(\number\nequation)}}

\def\label#1{\xdef #1{(\number\nequation)}
\ifnum\dflag = 1
{\escapechar = -1
\xdef\draftname{\littlefont\string#1}}
\fi}

\def\clabel#1#2{\xdef\eqlabel{(\number\nequation #2)}
\global\cflag = 1
\xdef #1{\eqlabel}
\ifnum\dflag = 1
{\escapechar = -1
\xdef\draftname{\string#1}}
\fi}

\def\cclabel#1#2{\xdef\eqlabel{#2)}
\global\cflag = 1
\xdef #1{\eqlabel}
\ifnum\dflag = 1
{\escapechar = -1
\xdef\draftname{\string#1}}
\fi}


\def\eeq{}

\def\eqnn #1\eeq{$$ #1 $$}

\def\eq #1\eeq{
\ifnum\dflag = 0
{\xdef\draftname{\ }}
\fi 
$$ #1
\eqno{\eqlabel \rlap{\ \draftname}} $$
\nexteqno}







\def\eqa #1\eeq{
\ifnum\dflag = 0
{\xdef\draftname{\ }}
\fi 
$$ \eqalignno{ #1 } $$
\global\cflag = 0}


\def\ie{{\it i.e.\/}}


\def\nci#1#2#3{{\it Nuovo Cimento} {\bf #1} (19#2) #3}
\def\npb#1#2#3{{\it Nucl.\ Phys.} {\bf B#1} (19#2) #3}
\def\plb#1#2#3{{\it Phys.\ Lett.} {\bf #1B} (19#2) #3}

\def\prd#1#2#3{{\it Phys.\ Rev.} {\bf D#1} (19#2) #3}

\def\prep#1#2#3{{\it Phys.\ Rep.} {\bf C#1} (19#2) #3}
\def\prl#1#2#3{{\it Phys.\ Rev.\ Lett.} {\bf #1} (19#2) #3}


\global\nulldelimiterspace = 0pt



\def\frac#1#2{{{#1} \over {#2}}\,}  



\def\Dsl{\hbox{/\kern-.6700em\it D}} 
\def\dsl{\hbox{/\kern-.5300em$\partial$}}
\def\pxpsl{\hbox{/\kern-.5600em$p$}}
\def\ssl{\hbox{/\kern-.5300em$s$}}
\def\epssl{\hbox{/\kern-.5100em$\epsilon$}}
\def\delsl{\hbox{/\kern-.6300em$\nabla$}}
\def\lxpsl{\hbox{/\kern-.4300em$l$}}
\def\elxpsl{\hbox{/\kern-.4500em$\ell$}}
\def\kxpsl{\hbox{/\kern-.5100em$k$}}
\def\qxpsl{\hbox{/\kern-.5000em$q$}}
\def\sla#1{\raise.15ex\hbox{$/$}\kern-.57em #1}



\def\roughly#1{\mathrel{\raise.3ex\hbox{$#1$\kern-.75em\lower1ex\hbox{$\sim$}}}}
\def\lsim{\roughly<}






\def\ssl{{\sss L}}







\barsoff


\def\ptran{p_{\sss T}}
\def\lft{{\sss L}}
\def\mx{M_{\sss X}}
\def\sigmahat{{\hat\sigma}}
\def\shat{\hat s}
\def\xs{X_{\sss S}}
\def\xv{X_{\sss V}}
\def\ks{k_{\sss S}}
\def\kv{k_{\sss V}}
\def\lntwoab{\ln\left[{2-\alpha-\beta}\over{2-\alpha+\beta}\right]}
\def\lnab{\ln\left[{\alpha+\beta}\over{\alpha-\beta}\right]}


\rightline{UdeM-LPN-TH-94-190}
\rightline{McGill-94/12}
\rightline{March 1994}
\vskip1truein

\centerline{\bf SINGLE DILEPTON PRODUCTION AT $e^+e^-$, $e\gamma$ AND
$\gamma\gamma$ COLLIDERS}

\vskip1.5truecm
\authors
\centerline{N. Lepor\'e,${}^a$ B. Thorndyke,${}^a$ H. Nadeau${}^b$ and D.
London${}^a$}
\vskip .15in
\centerline{\it ${}^a$ Laboratoire de Physique Nucl\'eaire, Universit\'e de
Montr\'eal}
\centerline{\it C.P. 6128, Montr\'eal, Qu\'ebec, CANADA, H3C 3J7.}
\vskip .1in
\centerline{\it ${}^b$ Physics Department, McGill University}
\centerline{\it 3600 University St., Montr\'eal, Qu\'ebec, CANADA, H3A 2T8.}
\endauthors

\abstract
We calculate the cross sections for the single production of doubly-charged
dileptons, both scalar and vector, at $e^+e^-$, $e\gamma$ and
$\gamma\gamma$ colliders at $\sqrt{s}=500$ GeV and 1 TeV. The $e\gamma$
mode is by far the most promising -- dileptons whose coupling is as weak as
$O(10^{-4})\alpha_{em}$ can be observed, for masses virtually up to the
kinematic limit.  Dileptons of mass up to $\sqrt{s}$ can also be seen in
$e^+e^-$ and $\gamma\gamma$ colliders, for couplings of order
$\alpha_{em}$. In all three colliders, most of the cross section comes from
events in which the only particles detected are $e^-e^-$, the decay
products of the dilepton, yielding an unmistakeable experimental signature.
\endabstract

\vfill\eject

\ref\frampton{P.H. Frampton and B.-H. Lee, \prl{64}{90}{619}.}
\ref\pisano{F. Pisano and V. Pleitez, \prd{46}{92}{410}; P.H. Frampton,
\prl{69}{92}{2889}, R. Foot, O.F. Hernandez, F. Pisano and V. Pleitez,
\prd{47}{93}{4158}.}
\ref\rizzo{T.G. Rizzo, \prd{45}{92}{42}.}
\ref\dilepcoll{T.G. Rizzo, \prd{46}{92}{910}; J. Agrawal, P.H. Frampton and
D. Ng, \npb{386}{92}{267}.}
Virtually all models of physics beyond the standard model (SM) predict the
existence of new, exotic particles. Over the past several years, an
enormous amount of effort has gone into exploring the possibilities for the
detection of such particles at present and future colliders. In one
interesting class of models, the gauge group for leptons is expanded from
the $SU(2)_\lft$ of the SM to $SU(3)$. This happens, for example, in
$SU(15)$ grand unified theories \frampton, or in models with an $SU(3)_\lft
\times U(1)$ gauge symmetry \pisano. In such models one finds dileptons
$(X^{--}, X^-)$, bosons (scalar or vector) which couple to pairs of
leptons. Their masses might very well be low enough that these particles
could be produced directly in the next generation of particle accelerators
\rizzo, \dilepcoll.

\ref\backscatter{I.F. Ginzburg, G.L. Kotkin, V.G. Serbo and V.I. Telnov,
Pis'ma ZhETF {\bf 34} (1981) 514; Sov.\ Yad.\ Fiz.\ {\bf 38} (1983) 372;
Nucl.\ Instr.\ Methods {\bf 205} (1983) 47; I.F. Ginzburg, G.L. Kotkin,
S.L. Panfil, V.G. Serbo and V.I. Telnov, Sov.\ Yad.\ Fiz.\ {\bf 38} (1983)
1021; Nucl.\ Instr.\ Methods {\bf 219} (1984) 5.}
Among future colliders, one of the most interesting is a high-energy linear
$e^+e^-$ collider. Not only does it provide a clean environment, but there
is also the possibility of backscattering laser light off one or both of
the beams \backscatter, creating an $e\gamma$ or $\gamma\gamma$ collider.

In this paper, we will examine the single production of doubly-charged
dileptons at such colliders. Rizzo has already taken a first step in this
calculation -- in Ref.~\rizzo\ he presents the cross sections for both
single and pair production of scalar dileptons at high-energy $e^+e^-$
colliders, noting that, for a large range of masses and couplings, single
production of dileptons dominates over pair production. We go beyond
Rizzo's work in three ways. First, we consider both scalar and vector
dileptons. Second, we compute the production cross section for all three
collider modes: $e^+e^-$, $e\gamma$ and $\gamma\gamma$. The third point is
a bit more technical -- these cross sections are infrared divergent in the
limit as $m_e\to 0$. We make an improvement to the procedure of regulating
this divergence, which results in an increase in the cross sections by a
factor of 3-5 relative to Rizzo's results. We will discuss this last point
in more detail below.

\ref\framng{P.H. Frampton and D. Ng, \prd{45}{92}{4240}.}
\ref\carlson{E.D. Carlson and P.H. Frampton, \plb{283}{92}{123}.}
\ref\dilepmass{H. Fujii, S. Nakamura and K. Sasaki, \plb{299}{93}{342}; H.
Fujii, Y. Mimura, K. Sasaki and T. Sasaki, TIT/HEP-234, BNL-49435,
YNU-HEPTh-93-103, hep-ph/9309287 (1993).}
\ref\studilep{K. Sasaki, \plb{308}{93}{297}.}
At present, the constraints on the dilepton mass are not very stringent
\framng-\dilepmass. The strongest bound on the mass of doubly-charged
vector dileptons comes from low-energy Bhabha scattering \framng:
$\mx/g_{3l} > 340$ GeV (95\% C.L.), while the mass of singly-charged vector
dileptons must obey $\mx/g_{3l} > 640$ GeV (90\% C.L.) from polarized muon
decay \carlson. Here, $g_{3l}$ is the $SU(3)$ gauge coupling constant,
which is expected to be of the same order of magnitude as the SM gauge
couplings. For example, $g_{3l}=1.19e$ for the $SU(15)$ GUT \frampton, and
$g_{3l}=2.07e$ in the $SU(3)_\lft \times U(1)$ model \pisano. Thus, the
doubly-charged dilepton mass must be greater than only $O(100)$ GeV. (Of
course, if the mass difference between the singly- and doubly-charged
dileptons is large, there will be significant constraints from the oblique
electroweak parameters \studilep.)

The Lagrangian describing the interactions of doubly-charged vector and
scalar dileptons is
\eq
{\cal L}_{\sss X} = -{g_{3l}\over \sqrt{2}} \, X_\mu^{++} e^{\sss T} C
\gamma^\mu \gamma_5 e + {g\over\sqrt{2}} \, X^{++} e^{\sss T} C
(1 - \lambda\gamma_5) e + h.c.
\eeq
In the above, the vector coupling of the vector dilepton vanishes by Fermi
statistics. Also, the coupling of the scalar dilepton is necessarily
chiral, since it couples to two left-handed or right-handed electrons
\rizzo, so that $\lambda=\pm 1$. In the following, we shall take the two
couplings, $g_{3l}$ and $g$, to be completely arbitrary, although, as
mentioned above, it should be remembered that $g_{3l}$ is expected to be
$\sim e$.

\ref\effphoton{V.M. Budnev, I.F. Ginzburg, G.V. Meledin and V.G. Serbo,
\prep{15}{75}{182}.}
\ref\efffermion{I.F. Ginzburg and V.G. Serbo, Novosibirsk preprint ``The
$\gamma\gamma\to Z l^+l^-$ and $\gamma\gamma\to Z q{\overline q}$ processes
at the polarized photon beams'' (unpublished). See also P. Kessler,
\nci{17}{60}{809}; V.N. Baier, V.S. Fadin and V.A. Khoze,
\npb{256}{85}{189}.}
The diagrams which lead to single dilepton production in $e\gamma$
colliders are shown in Fig.~1. These are also the diagrams which dominate
single dilepton production in $e^+e^-$ and $\gamma\gamma$ colliders. In
$e^+e^-$ colliders, an energetic virtual photon is emitted from the $e^+$
beam, leading to the $e^-\gamma$ collisions of Fig.~1. The total cross
section is then obtained by using the effective photon approximation. That
is, the cross section for the process $e^-\gamma \to X^{--}e^+$ is folded
together with the photon distribution function to give $\sigma_{ee}(s)$,
the cross section for the process $e^+e^- \to X^{--} e^+ e^+$ at
centre-of-mass energy $\sqrt{s}$:
\label\sigmaee
\eq
\sigma_{ee}(s) = \int_{s_{th}/s}^1 d\tau f_\gamma(\tau) \sigmahat(\tau s).
\eeq
Here, $\sigmahat(\tau s)$ is the cross section for the sub-process
$e^-\gamma \to X^{--}e^+$ at centre-of-mass energy $\sqrt{\shat}$, with
$\shat = \tau s$, and $\sqrt{s_{th}}$ is the threshold energy for the
production of the final states in this process. The photon distribution
function $f_{\gamma}(\tau)$ is given by \effphoton:
\label\photdist
\eq
f_\gamma(\tau) = {\alpha_{em}\over 2\pi} \left\{
{ \left[1+(1-\tau)^2 \right] \over \tau}
\ln \left[{s\over 4m_e^2}{(1-2\tau+\tau^2)\over (1-\tau+\tau^2/4)}\right]
+ \tau \ln \left( {2-\tau\over\tau} \right)
+ {2(\tau-1)\over \tau} \right\}~.
\eeq
Similarly, in $\gamma\gamma$ colliders, one of the photons turns into a
real $e^+$ (which is soft) and a virtual energetic $e^-$. The total cross
section is then obtained by using the effective fermion approximation:
\label\sigmagg
\eq
\sigma_{\gamma\gamma}(s) = 2\int_{s_{th}/s}^1 d\tau f_e(\tau)
\sigmahat(\tau s),
\eeq
in which  $\sigma_{\gamma\gamma}(s)$ is the total cross section for the
process $\gamma\gamma \to X^{--} e^+ e^+$ at centre-of-mass energy
$\sqrt{s}$, and a factor of 2 has been included since the $e^-$ can be
emitted from either photon. The effective fermion function is given by
\efffermion\
\label\fermiondist
\eq
f_e(\tau) = {\alpha_{em}\over 2\pi} \left\{ \left[\tau^2+(1-\tau)^2 \right]
\ln \left[{s\over 4m_e^2} (1-\tau)^2 \right] + 2\tau(1-\tau) \right\}~.
\eeq
Note that there are other diagrams leading to single dilepton production in
both $e^+e^-$ and $\gamma\gamma$ colliders. However, the presence of the
term $\ln(s/4m_e^2)$ in both the effective photon and effective fermion
functions leads to a large enhancement of the contributions of the diagrams
in Fig.~1 relative to these other diagrams. The error incurred by
neglecting the other diagrams is estimated to be only about 5\%
\efffermion.

\ref\lq{This procedure was also used in the calculation of single
leptoquark production at $e^+e^-$ and $\gamma\gamma$ colliders, see G.
B\'elanger, D. London and H. Nadeau, UdeM-LPN-TH-93-152, McGill-93/23, to
be published in Physical Review {\bf D}.}
{}From the above discussion, it is clear that the most important step in
the calculation is the computation of the diagrams in Fig.~1. Before
presenting the results, let us first address a key detail. If the mass of
the electron is neglected, the second diagram in Fig.~1 diverges. This
happens in that region of phase space in which the 3-momentum of the final
$e^+$ is parallel to that of the initial photon. One way to deal with this
is to impose a $\ptran$ cut of, say, 10 GeV on the final electron \rizzo.
While this solves the problem, a large fraction of the total cross section
is eliminated in the process. An alternative procedure, which is the one
used in this paper, is to use the nonzero electron mass as the regulator
\lq. (In this case, $s_{th} = (\mx + m_e)^2$ in Eqs.~\sigmaee\ and
\sigmagg.) As we will see, this results in a substantial increase in the
cross section compared to the $\ptran$-cut procedure. Note that, regardless
of the collider mode ($e^+e^-$, $e\gamma$ or $\gamma\gamma$), this increase
corresponds to the inclusion of events which are completely collinear, \ie\
events in which the $X^{--}$ and the other final-state particle(s) go down
the beam pipe. However, this does not create any experimental problems.
Since the dilepton will then decay to $e^-e^-$, as far as the detector is
concerned the process is effectively $e^+e^-~(or~e\gamma~or~\gamma\gamma)
\to e^-e^-$, giving a signal which is unmistakable, and has, of course,
virtually no SM background.

The cross sections for single dilepton production at $e^+e^-$, $e\gamma$
and $\gamma\gamma$ colliders are presented for two values of the
centre-of-mass energy, $\sqrt{s}=500$ GeV and 1 TeV. We assume the
integrated luminosity at all 3 colliders to be 10 ${\rm fb}^{-1}$ at 500
GeV, and 60 ${\rm fb}^{-1}$ at 1 TeV. As a figure of merit, we require 25
events for discovery, which corresponds to a cross section of 2.5 $fb$ at
$\sqrt{s}=500$ GeV and 0.4 $fb$ at 1 TeV. Note that we consider {\it only}
$X^{--}$ production. If one includes both $X^{--}$ and $X^{++}$ production,
the cross sections must be multiplied by a factor of 2.

We focus first on the scalar dileptons, $\xs$. We parametrize the strength
of the dilepton coupling by comparing it to the electromagnetic
interaction, \ie\ $g^2=4\pi \ks\alpha_{em}$, and allowing $\ks$ to vary.
The cross section for $e^-\gamma\to \xs^{--}e^+$ is found to be
\label\sigmascalar
\eq
\eqalign{
\sigma_{\sss S} (s) & = {\pi \ks \alpha_{em}^2 \over s^2} \left[ \beta
\left( {3\over 2}s + {17\over 2}\mx^2 \right) + 8 \mx^2 \lntwoab \right.
\cr
& \left. \qquad\qquad\qquad\qquad\qquad + ~{ \left(s^2 - 2\mx^2 (s - \mx^2)
\right) \over s} \lnab \right], \cr}
\eeq
independent of the chirality of the scalar dilepton coupling. In the above,
\label\alphadef
\eq
\alpha\equiv 1 - {(\mx^2 - m_e^2)\over s}
\eeq
and
\label\betadef
\eq
\beta \equiv\left( 1 - 2 \, {(\mx^2+m_e^2)\over s}
+ {(\mx^2-m_e^2)^2\over s^2} \right)^{1\over 2}~.
\eeq
The above cross section is written in such a way that it is clear that
$\sigma_{\sss S} (s) \to 0$ as the kinematic limit $\beta \to 0$ is
reached. Away from the kinematic limit, the logarithms in Eq.~\sigmascalar\
can be written in the more transparent forms:
\label\simplelogs
\eq
\eqalign{
\lnab & \to \ln\left[{ (s-\mx^2)^2 \over s \, m_e^2 }\right], \cr
\lntwoab & \to \ln\left[{\mx^2\over s}\right].\cr}
\eeq
A comparison of the two regulation procedures --- a nonzero $m_e$ and a 10
GeV $\ptran$ cut --- is straightforward. Eq.~\sigmascalar\ still holds when
a $\ptran$ cut is used, with (i) the replacement of $m_e$ in
Eq.~\simplelogs\ by $\ptran$ and the neglect of $m_e$ everywhere else, and
(ii) the addition of a single finite piece,
$(\pi\ks\alpha_{em}^2/s^2)\beta(s+\mx^2)$. This agrees with the results of
Ref.~\rizzo. By comparing $\ln(s/m_e^2)$ with $\ln(s/\ptran^2)$ (these are
typically the largest contributions), one sees that one gains a factor of
roughly 3-5 by using $m_e$ as a regulator. This expectation is borne out
quantitatively, as we will see below.

The cross section for the process $e^-\gamma\to \xs^{--}e^+$ with dilepton
coupling strength $\ks=1$ is shown in Fig.~2, as a function of the dilepton
mass, $\mx$. At both $\sqrt{s}=500$ GeV and 1 TeV, for virtually the entire
range of $\mx$, the production cross section is orders of magnitude above
the cross sections required for discovery (2.5 $fb$ at $\sqrt{s}=500$ GeV,
0.4 $fb$ at 1 TeV). In other words, for $\ks=1$, dileptons with masses
essentially up to the kinematic limit will be easily observable. Since the
cross section is linear in $\ks$, it is straightforward to scale the
results shown in Fig.~2 to other values of $\ks$. This shows that scalar
dileptons of $\mx\lsim\sqrt{s}$ with couplings as small as
$\ks=5{\hbox{-}}7\times 10^{-4}$ can be seen in high-energy $e\gamma$
collisions.

In Fig.~3 we present the cross sections for the process $e^+e^-\to \xs^{--}
e^+e^+$, for three values of the coupling, $\ks=1$, 0.1 and 0.01. To
calculate these cross sections, we use the effective photon approximation
described by Eq.~\sigmaee\ above. In Figs.~3a and 3b the straight line
corresponds to the assumed discovery cross section. Thus, for example, at
$\sqrt{s}=1$ TeV, dileptons with coupling strength $\ks=1$ (or $\ks=0.1$ or
.01) can be seen for $\mx \lsim 990$ GeV (or $\mx\lsim 950$ GeV or 700
GeV). We have also calculated these cross sections using a $\ptran$ cut to
regulate the forward divergence, and we reproduce the results of Rizzo
\rizzo. By comparing the results shown in Fig.~3 with those in Ref.~\rizzo\
(remembering that in Ref.~\rizzo\ there is an additional factor of 2 due to
both $\xs^{--}$ and $\xs^{++}$ production), we see that the procedure of
using a nonzero $m_e$ as a regulator does indeed increase the cross section
by a substantial factor compared to the $\ptran$-cut procedure.

Finally, Fig.~4 shows the cross sections for $\gamma\gamma\to \xs^{--}
e^+e^+$, for $\ks=1$, 0.1 and 0.01. These cross sections have been
calculated using the effective fermion approximation (Eq.~\sigmagg).
Comparing Fig.~4 to Fig.~3, we note that, for most values of $\mx$, single
dilepton production in $e^+e^-$ collisions is greater than that in
$\gamma\gamma$ collisions. It is only for values of $\mx$ quite close to
$\sqrt{s}$ that the cross section for $\gamma\gamma\to \xs^{--} e^+e^+$
exceeds that for $e^+e^-\to \xs^{--} e^+e^+$. However, when one takes into
account that the luminosity of a $\gamma\gamma$ collider is at most 80\% of
the parent $e^+e^-$ collider, one concludes that the $e^+e^-$ mode is
better than the $\gamma\gamma$ mode for single dilepton production. (Of
course, neither can compete with the $e\gamma$ mode, as is clear from
Fig.~2.)

We now turn to vector dileptons, $\xv$. As was done in the case of scalar
dileptons, we parametrize the strength of the coupling as $g_{3l}^2=4\pi
\kv\alpha_{em}$. (However, it should be remembered that, since $g_{3l}$ is
a gauge coupling, $\kv$ is expected to be $\sim 1$.) The cross section for
$e^-\gamma\to \xv^{--}e^+$ is given by
\label\sigmavector
\eq
\eqalign{
\sigma_{\sss V} (s) & = {\pi \kv \alpha_{em}^2 \over s} \left\{ \beta
\left( 2 + {8 s \over \mx^2} + {13\over 2}\,{\mx^2\over s} \right)
+ {(s^2-2s\mx^2 +2\mx^4) \over s^2} \lnab \right. \cr
& \left. \qquad\qquad\qquad\qquad + { \left(-s^3 + 12 s^2 \mx^2 + 18
s \mx^4 - 4 \mx^6\right) \over 2 s^2 \mx^2} \lntwoab \right\}, \cr}
\eeq
where $\alpha$ and $\beta$ are defined in Eqs.~\alphadef\ and \betadef.

The cross section for $e^-\gamma\to \xv^{--}e^+$ with $\kv=1$ is shown in
Fig.~5. As was the case for scalar dileptons, at both $\sqrt{s}=500$ GeV
and 1 TeV the cross sections are enormous, so that dileptons with masses
almost up to the kinematic limit are easily observable. Indeed, vector
dileptons with couplings as small as $\kv=3{\hbox{-}}4\times 10^{-4}$ can
be seen in $e^-\gamma\to \xv^{--}e^+$.

In Figs.~6 and 7 we present the cross sections for the processes $e^+e^-\to
\xv^{--}e^+e^+$ and $\gamma\gamma\to \xv^{--}e^+e^+$, respectively, for
three values of the coupling, $\kv=1, 0.1$ and 0.01. In both processes, for
$\kv\sim 1$, which is favoured, dileptons with masses virtually up to the
kinematic limit are observable. As was the case for scalar dileptons, a
comparison of Figs.~6 and 7 reveals that the production cross section in
$e^+e^-$ mode is greater than that in $\gamma\gamma$ mode for most values
of $\mx$. Therefore, $e^+e^-$ collisions are better than $\gamma\gamma$
collisions for producing single vector dileptons.

To summarize, we have calculated the single production of dileptons, both
scalar and vector, in $e^+e^-$, $e\gamma$ and $\gamma\gamma$ colliders at
$\sqrt{s}=500$ GeV and 1 TeV. In $e^+e^-$ and $\gamma\gamma$ collisions,
the contribution from the subprocess $e^-\gamma\to X^{--}e^+$ dominates the
production cross section. We have used the nonzero electron mass to
regulate the infrared divergence in that region of phase space in which the
momenta of the final $e^+$ and the initial $\gamma$ are parallel. This
results in an increase in the cross section by a factor of 3-5 compared to
the method of putting a $\ptran$ cut on the final $e^+$. With this method,
most of the cross section comes from events which go down the beam pipe.
However, this causes no problems -- when the dilepton decays, this results
in an unmistakeable signal in the detector:
$e^+e^-~(or~e^-\gamma~or~\gamma\gamma) \to e^-e^-$, which is virtually
background-free.

Of the three colliders, the $e\gamma$ mode is by far the most promising --
scalar and vector dileptons with masses up to essentially $\sqrt{s}$, and
whose coupling strength is equal to $\alpha_{em}$, will be copiously
produced in the process $e^-\gamma\to X^{--}e^+$. In fact, it will be
possible to detect dileptons whose coupling strength is as small as
$O(10^{-4})\alpha_{em}$. As for $e^+e^-$ and $\gamma\gamma$ colliders, even
though the cross sections for single dilepton production are much smaller
here than they are in the $e\gamma$ collider, it is still possible to
detect dileptons with masses up to the kinematic limit, for a dilepton
coupling strength equal to $\alpha_{em}$. Even dileptons whose coupling is
weaker are detectable over a large range of masses. It should also be noted
that, for a given dilepton mass, the production cross section is greater in
$e^+e^-$ colliders than in $\gamma\gamma$ colliders.

\bigskip
\centerline{\bf Acknowledgments}
\medskip
\noindent
This research was partially funded by the N.S.E.R.C.\ of Canada and les
Fonds F.C.A.R.\ du Qu\'ebec.

\vfill\eject
\centerline{Figure Captions}
\bigskip

\topic{Figure (1)} The three diagrams contributing to the process
$e^-\gamma \to X^{--}e^+$.

\topic{Figure (2)} Cross section for the process $e^-\gamma\to \xs^{--}e^+$
at (a) $\sqrt{s}=500$ GeV, and (b) $\sqrt{s}=1$ TeV, for $\ks=1$.

\topic{Figure (3)} Cross section for the process $e^+e^-\to \xs^{--}e^+e^+$
at (a) $\sqrt{s}=500$ GeV, and (b) $\sqrt{s}=1$ TeV, for $\ks=1$ (solid
line), $\ks = 0.1$ (dash-dot line) and $\ks=0.01$ (dashed line). The
horizontal line is the (assumed) discovery cross section of 2.5 $fb$ (0.4
$fb$) at $\sqrt{s}=500$ GeV (1 TeV).

\topic{Figure (4)} Cross section for the process $\gamma\gamma\to
\xs^{--}e^+e^+$ at (a) $\sqrt{s}=500$ GeV, and (b) $\sqrt{s}=1$ TeV, for
$\ks=1$ (solid line), $\ks = 0.1$ (dash-dot line) and $\ks=0.01$ (dashed
line). The horizontal line is the (assumed) discovery cross section of 2.5
$fb$ (0.4 $fb$) at $\sqrt{s}=500$ GeV (1 TeV).

\topic{Figure (5)} Cross section for the process $e^-\gamma\to \xv^{--}e^+$
at (a) $\sqrt{s}=500$ GeV, and (b) $\sqrt{s}=1$ TeV, for $\kv=1$.

\topic{Figure (6)} Cross section for the process $e^+e^-\to \xv^{--}e^+e^+$
at (a) $\sqrt{s}=500$ GeV, and (b) $\sqrt{s}=1$ TeV, for $\kv=1$ (solid
line), $\kv = 0.1$ (dash-dot line) and $\kv=0.01$ (dashed line). The
horizontal line is the (assumed) discovery cross section of 2.5 $fb$ (0.4
$fb$) at $\sqrt{s}=500$ GeV (1 TeV).

\topic{Figure (7)} Cross section for the process $\gamma\gamma\to
\xv^{--}e^+e^+$ at (a) $\sqrt{s}=500$ GeV, and (b) $\sqrt{s}=1$ TeV, for
$\kv=1$ (solid line), $\kv = 0.1$ (dash-dot line) and $\kv=0.01$ (dashed
line). The horizontal line is the (assumed) discovery cross section of 2.5
$fb$ (0.4 $fb$) at $\sqrt{s}=500$ GeV (1 TeV).
\endtopic

\listrefs

\bye